\documentclass[11pt]{article}
\usepackage{amsmath}
\usepackage{amsthm}
\usepackage{amsfonts}
\usepackage{bbm}
\usepackage{graphicx}
\usepackage{sectsty}
\RequirePackage{setspace}
\usepackage{appendix}
\usepackage{float}
\usepackage{url}
\usepackage{caption}
\usepackage{lineno}
\usepackage[round]{natbib}
\usepackage[table,xcdraw]{xcolor}
\usepackage{float}

\newcommand{\captionfonts}{\footnotesize}
\makeatletter 
\long\def\@makecaption#1#2{%
\vskip\abovecaptionskip
\sbox\@tempboxa{{\captionfonts #1: #2}}%
\ifdim \wd\@tempboxa >\hsize
{\captionfonts #1: #2\par}
\else
\hbox to\hsize{\hfil\box\@tempboxa\hfil}%
\fi
\vskip\belowcaptionskip}
\makeatother 

\begin{document}
\title{{\bf Quantum entanglement partly demystified}}

\author{Diederik Aerts and Massimiliano Sassoli de Bianchi\vspace{0.5 cm} \\ 
\normalsize\itshape
Center Leo Apostel for Interdisciplinary Studies, \\ \itshape Brussels Free University, 1050 Brussels, Belgium\vspace{0.5 cm} \\ 
\normalsize
E-Mails: \url{diraerts@vub.ac.be}, \ \url{msassoli@vub.ac.be}
}

\date{}

\maketitle

\begin{abstract}
\noindent
We consider a simple string model to explain and partly demystify the phenomenon of quantum entanglement. The model in question has nothing to do with string theory: it uses macroscopic strings that can be acted upon by Alice and Bob in ways that violate, or fail to violate, in different ways Bell-CHSH inequalities and the no-signaling conditions, also called marginal laws. We present several variants of the model, to address different objections that may arise. This allows us to make fully visible what the quantum formalism already suggests, about the nature of the correlations associated with entangled states, which appear to be created in a contextual manner at each execution of a joint measurement. We also briefly present the hidden measurement interpretation, whose rationale is compatible with the mechanism suggested by our string model, then offer some final thoughts about the possibility that the quantum entanglement phenomenon might affect not only states, but also measurements, and that our physical reality would be predominantly non-spatial in nature.
\end{abstract}

\section{Introduction\label{intro}}

Schr\"{o}dinger first introduced the term \emph{entanglement} in the 1930s, and he believed it was the characteristic trait that separated quantum mechanics from classical mechanics \citep{Schrodinger1935}. Thirty years later, John Bell's famous inequalities were able to test the presence of entanglement in bipartite systems \citep{Bell1964}. Despite Bell's belief that his inequalities would not be violated, the predictions of quantum theory are no longer put into question today. Quantum entanglement has been shown to be preservable over very large spatial distances \citep{aspect1982a,aspect1983,aspect1982b,tittel1998,weihs1998,giustina2013,christensen2013,hensen-etal2016}, and the debate about the profound meaning of the phenomenon has never stopped.

In this article, we show that it is possible to partially demystify entanglement by explaining it in a way that is perfectly compatible with what the quantum formalism already suggests. We will do so by following the approach initiated by one of us in the 1980s, which started from the analysis of simple but sophisticatedly defined macroscopic systems, able to partially imitate the behavior of the microscopic ones, shedding light on the quantum measurement problem in general and on the correlations emerging from quantum entangled systems in particular \citep{Aerts1982,a1984,Aerts1986,Aerts1990,Aerts1991}. These ideas were more recently incorporated in a rather general representation of quantum states and measurements, called the \emph{extended Bloch representation of quantum mechanics} \citep{AertsSassolideBianchi2014,AertsSassoli2016}, which allows to explain quantum measurement and quantum entanglement in terms of ``hidden'' (measurement) interactions, responsible for the reduction of the system's state. This is to say that although in this article we will be dealing with very simple toy model systems, the perspective our analysis opens up is broad and still under investigation. 

Another important aspect of what we will explain is its didactical value. Indeed, if the ideas behind the simple models that we will describe are not new, although to our knowledge what we have called ``Variant 4'' of the model has never been described to date, it is the way in which we will present them that is new and, we hope, capable of capturing the attention even of those who, until now, have not given sufficient importance to the clarification that these models allow especially considering that they also find a more general representation within the quantum formalism and the already mentioned extended Bloch representation of quantum mechanics. 

More precisely, starting from a very simple situation, which demonstrates the possibility for a macroscopic system to maximally violate the Bell-CHSH inequalities, hence revealing a possible mechanism behind the quantum correlations, we proceed by presenting a first possible objection. This will allow us to propose a first variant of the model, which will give rise to a subsequent objection, and so on, up to variant number 4. At that point, the objection will be that these simple systems, however clarifying, may have nothing to do with what happens at the microscopic level. And here we come to the last part of this article, where we briefly present the gist of the extended Bloch representation and the associated hidden measurement interpretation. 

In the Conclusion section, we address a last objection, that of the absence of a connective structure detectable in space, able to connect the entangled entities and explain their behavior as if they were a single interconnected whole. Here our analysis comes into contact with the real mystery that the phenomenon of entanglement reveals, which is not possible to demystify and which leads us to contemplate a physical reality of a non-spatial nature, where the Euclidean or Mikowski spaces are simply seen as theaters capable of representing the relationships between the different macroscopic entities, and not as a background canvas for all of reality. Finally, we will also evoke the possibility that entanglement could additionally manifest at the level of the measurements, and not only of the states \citep{aertsetal2019}.

\section{Bell-CHSH inequalities}

We consider a \emph{bipartite system}, such that it is possible to identify two parts of it, interpretable as two possibly interconnected sub-systems, forming the whole system. These two sub-systems may be more or less easy to characterize, in the sense that it may not be always clear where one sub-system ends and the other begins. What is important, however, is that the ability to distinguish Alice's actions, on one sub-system, from Bob's actions, on the other sub-system, is not lost, Alice and Bob being here the two fictional characters traditionally used to describe these actions. 

The paradigmatic example of a quantum bipartite system in an entangled state is that proposed long ago by David Bohm, of two spin-${1\over 2}$ fermionic entities in a rotationally invariant singlet state \citep{bohm1951}:
\begin{equation}
|s\rangle={1\over \sqrt{2}}\left(|+\rangle\otimes|-\rangle -|-\rangle\otimes|+\rangle\right)
\label{singletstate2}
\end{equation}
The universally used test for the presence of entanglement is that provided by the Clauser Horne Shimony Holt (CHSH) version of Bell’s inequalities \citep{Clauser1969}. They can be formulated by considering four joint measurements, which we will denote $AB$, $AB'$, $A'B$ and $A'B'$, where $A$ and $A'$ are the two measurements Alice can freely select and execute on her sub-system, and $B$ and $B'$ are the two measurements that Bob can freely select and execute on his sub-system. 

In the case of a two-spin composite system, the joint measurement $AB$ corresponds to the situation where Alice measures one of the two spins, with a Stern-Gerlach oriented along the $A$-axis, whereas Bob measures the other spin, using a Stern-Gerlach oriented along the $B$-axis, and same for the other three joint measurements, which use Stern-Gerlach apparatuses oriented along the $A'$ and $B'$ axes, respectively. One then introduces the four linear combinations: 
\begin{eqnarray}
&&A_{\rm CHSH}\equiv -E_{AB}+E_{AB'}+E_{A'B}+E_{A'B'}\nonumber\\ 
&&B_{\rm CHSH}\equiv \phantom{-} E_{AB}-E_{AB'}+E_{A'B}+E_{A'B'}\nonumber\\
&&C_{\rm CHSH}\equiv \phantom{-}E_{AB}+E_{AB'}-E_{A'B}+E_{A'B'}\nonumber\\
&&D_{\rm CHSH}\equiv \phantom{-}E_{AB}+E_{AB'}+E_{A'B}-E_{A'B'}
\label{CHSH-quantity}
\end{eqnarray}
where we have defined the \emph{correlation function}: 
\begin{equation}
E_{AB}=(P_{AB}^{++}+P_{AB}^{--})-(P_{AB}^{+-}+P_{AB}^{-+})
\label{correlations}
\end{equation}
which accounts for the ``$++$'' and ``$--$'' correlated outcomes with a positive sign and for the ``$+-$'' and ``$-+$'' anticorrelated outcomes with a negative sign, where $P_{AB}^{ij}$ is the probability that the joint measurement $AB$ gives the outcome $i$ for $A$ and $j$ for $B$, with $i,j\in\{+,-\}$. And of course the correlation functions $E_{A'B}$, $E_{AB'}$ and $E_{A'B'}$ are defined in a similar way. Bell-CHSH inequalities then correspond to the four inequalities:
\begin{equation}
-2\leq A_{\rm CHSH},B_{\rm CHSH},C_{\rm CHSH},D_{\rm CHSH}\leq 2
\label{bounds}
\end{equation}

We will not discuss here in detail the case of two spin-$1/2$ entities in the singlet state (\ref{singletstate2}), since its treatment can be found in any good quantum physics textbook. Let us just mention that if $\alpha={\pi\over 4}$ is the angle between the $A$ and $B$ axes, and one additionally considers an angle of ${3\pi\over 4}$ between the $A$ and $B'$ axes, and an angle of ${\pi\over 4}$ between the $B$ and $A'$ axes, with angles of ${\pi\over 2}$ between $A$ and $A'$ and between $B$ and $B'$, one finds the values: $A_{\rm CHSH}=C_{\rm CHSH}=D_{\rm CHSH}=0$ and $B_{\rm CHSH}=-2\sqrt{2}$. Since $-2\sqrt{2}<-2$, inequality (\ref{bounds}) for $B_{\rm CHSH}$ is clearly violated, and corresponds to what is known as \emph{Cirel'son's bound}, the maximal violation one can achieve in standard quantum mechanics \citep{cirelson1980}, for as long as all joint measurements are described as product measurements relative to the same tensorial representation \citep{aertsetal2019}.

\section{A string model}

We consider a simple macroscopic string model that can violate (\ref{bounds}), inspired by the original ``vessels of water model'' \citep{Aerts1982} and from similar models usually described in terms of breakable elastic structures; see for instance \citep{Sassoli2013} and the references cited therein. After explaining why the model is able to do so, and why it could be illustrative of what also happens in microphysical systems, we consider a first possible objections, which will lead us to analyze a second variant of the model, then a second objection will come, leading to a third version of the model, then a fourth.

\subsection{Variant 1: white string}
\label{version 1}

The system on which Alice and Bob perform their joint experiments is a \emph{white string} of length $L$, made of a breakable material. It can be considered as a bipartite system as one can easily identify two parts of it, which are the two ends of the string on which Alice and Bob can act independently from one another. 

\begin{figure}[!ht]
\centering
\includegraphics[scale =0.24]{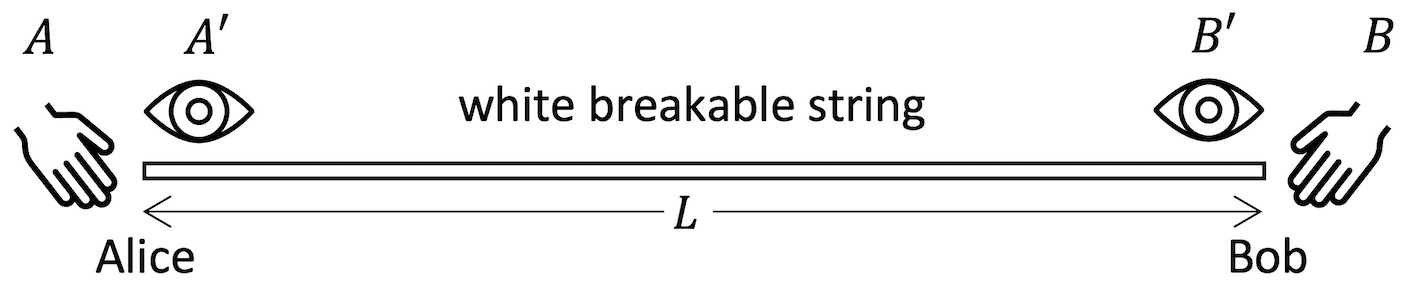}
\caption{Variant 1: a schematic representation of the four experiments $A$, $A'$, $B$ and $B'$ that Alice and Bob can jointly perform on a white string of length $L$.}
\label{Version1}
\end{figure}

Alice's experiments $A$ and $A'$ consists in measuring the length and color of her string fragment, respectively. More precisely, in experiment $A$ she pulls hard on her end of the string, to then measure the length of the collected fragment, for example using a yardstick. If the length is greater than $L/2$, the outcome is noted ``$+$,'' and ``$-$'' otherwise. Experiment $A'$, on the other hand, simply consists in observing the color of the string (without pulling it), and if she finds it is white, the outcome is noted ``$+$,'' and ``$-$'' otherwise. Bob's experiments $B$ and $B'$ are the same, but executed on his end side of the string. See the schematic representation in Figure~\ref{Version1}.

To calculate the correlation functions, one needs to determine the values of the joint probabilities. Here we assume that the string is uniform, hence it can break in any point with equal probability, when jointly pulled by Alice and Bob. This means that the probability that Alice collects a fragment of length greater than $L/2$, when jointly performing with Bob measurement $AB$, is simply $1/2$, and of course the same is true for Bob. Also, we assume that Alice always has enough time to see the color of the string, before it is pulled by Bob, when they preform $A'B$, and same for Bob when they perform $AB'$. And since the string is white, the probability to observe a non-white color is equal to zero. Also, in joint measurement $AB'$, Alice will pull the entire string, hence the observed length will be with certainty greater than $L/2$, and same for Bob in measurement $A'B$. Based on these observations, it is easy to convince oneself that we have the joint probabilities described in Table~\ref{t1}. 
\begin{table}[H]
\begin{center}
\begin{tabular}{|
>{\columncolor[HTML]{EFEFEF}}c |c|c|c|c|}
\hline
\cellcolor[HTML]{C0C0C0} & \cellcolor[HTML]{EFEFEF} $P^{++}$ & \cellcolor[HTML]{EFEFEF} $P^{+-}$& \cellcolor[HTML]{EFEFEF} $P^{-+}$& \cellcolor[HTML]{EFEFEF} $P^{--}$\\ \hline
$AB$ & $0$ & $1/2$ & $1/2$ & $0$ \\ \hline
$AB'$ & $1$ & $0$ & $0$ & $0$ \\ \hline
$A'B$ & $1$ & $0$ & $0$ & $0$ \\ \hline
$A'B'$ & $1$ & $0$ & $0$ & $0$ \\ \hline
\end{tabular}
\end{center}
\caption{Variant 1: the $16$ joint probabilities associated with the outcomes of Alice's and Bob's joint measurements on a white string, producing the value $A_{\rm CHSH}=4$, which maximally violates Bell-CHSH inequalities (\ref{bounds}).\label{t1}}
\end{table}

We see that joint measurement $AB$ produces perfect anticorrelations, hence $E_{AB}=-1$, whereas the other three joint measurements, $AB'$, $A'B$ and $A'B'$, produce perfect correlations, hence $E_{AB'}=E_{A'B}=E_{A'B'}=1$. Therefore, $B_{\rm CHSH}=C_{\rm CHSH}=D_{\rm CHSH}=0$, $A_{\rm CHSH}=4$, and we obtain an algebraically maximal violation of the CHSH inequalities (\ref{bounds}).

\subsection{Explanation of the model and first objection}

Some readers may be surprised to learn that a macroscopic system is capable of violating (\ref{bounds}), since there is still a widespread prejudice that Bell's inequalities have solely to do with the description of microscopic systems. Bell's inequalities, however, do not demarcate between microscopic and macroscopic systems, but between \emph{correlations of the first kind} and \emph{correlations of the second kind} \citep{Aerts1990}. More precisely, correlations that are contextually created by a joint measurement are called `of the second kind', whereas correlations that were already actual prior to the execution of a joint measurement, hence are not created by the latter, they are called `of the first kind'. 

When Alice and Bob perform the joint measurement $AB$, i.e., when they jointly pull their respective ends of the string, the latter will break in a point that cannot be predicted in advance. But by conservation of the matter with which the string is formed, the length of Alice's fragment will always be perfectly anticorrelated with that of Bob, as is clear that if $L_A$ is the length obtained by Alice and $L_B$ is the length obtained by Bob, we always necessarily have $L_A+L_B=L$, hence, if $L_A>L/2$, then $L_B<L/2$, and if $L_B>L/2$, then $L_A<L/2$. However, since the two lengths $L_A$ and $L_B$ do not pre-exist the joint measurement $AB$, we are in a situation where the latter creates the correlations, in the sense that it each time actualizes one among an infinite number of \emph{potential correlations}. In other words, they are correlations of the second kind.

\begin{figure}[!ht]
\centering
\includegraphics[scale =0.24]{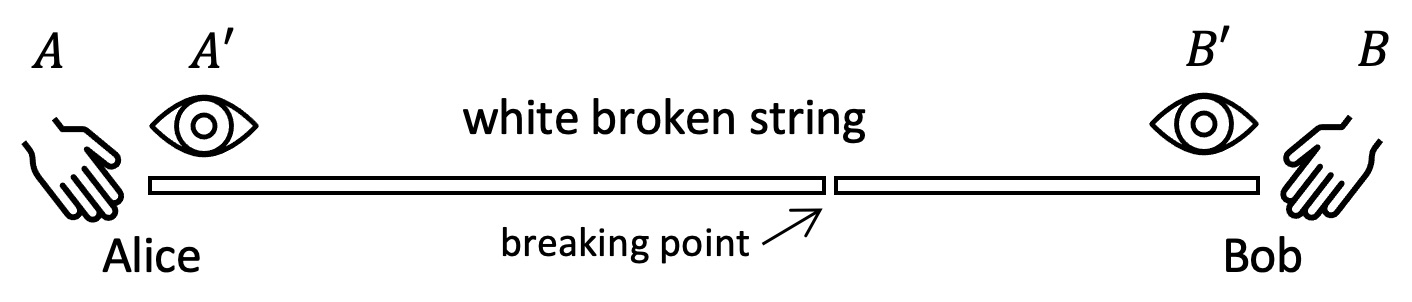}
\caption{Variant 1 with broken string: a schematic representation of the four experiments $A$, $A'$, $B$ and $B'$ that Alice and Bob can jointly perform on a white string of length $L$, in the situation where the string is already broken, here in a point such that Alice's fragment is longer than Bob's fragment.}
\label{Version1b}
\end{figure}

The model also contains correlations of the first kind, associated with the other three joint measurements, but not all the correlations need to be of the second kind to obtain a violation. It is sufficient that only some of them are. But if none of them are, then the violation of (\ref{bounds}) becomes impossible. To enable the reader to appreciate the difference between having or not having correlations of the second kind, we can consider the situation where a colleague of Alice and Bob, who likes to play pranks, has pre-cut the string at some unspecified point; see schematic representation in Figure~\ref{Version1b}. This means that the two lengths $L_A$ and $L_B$ are now actual before the execution of the joint measurements, hence, they are not anymore created by them, but only discovered.
The joint probabilities are now given in Table~\ref{t2}, and the difference from the previous situation is that when Alice executes measurement $A$, and Bob jointly executes measurement $B'$, Alice will not anymore pull the entire string, but only a fragment of it, and same for Bob when he executes $B$ and Alice executes $A'$.
\begin{table}[H]
\begin{center}
\begin{tabular}{|
>{\columncolor[HTML]{EFEFEF}}c |c|c|c|c|}
\hline
\cellcolor[HTML]{C0C0C0} & \cellcolor[HTML]{EFEFEF} $P^{++}$ & \cellcolor[HTML]{EFEFEF} $P^{+-}$& \cellcolor[HTML]{EFEFEF} $P^{-+}$& \cellcolor[HTML]{EFEFEF} $P^{--}$\\ \hline
$AB$ & $0$ & $1/2$ & $1/2$ & $0$ \\ \hline
$AB'$ & $1/2$ & $0$ & $1/2$ & $0$ \\ \hline
$A'B$ & $1/2$ & $1/2$ & $0$ & $0$ \\ \hline
$A'B'$ & $1$ & $0$ & $0$ & $0$ \\ \hline
\end{tabular}
\end{center}
\caption{The $16$ joint probabilities associated with the outcomes of Alice's and Bob's joint measurements on a white string, when the string is already broken, which do not anymore violate Bell-CHSH inequalities (\ref{bounds}).\label{t2}}
\end{table}

We now have $E_{AB'}=E_{A'B}=0$, and $E_{A'B'}=-E_{AB}=1$, implying that $B_{\rm CHSH}=C_{\rm CHSH}=0$ and $A_{\rm CHSH}=-D_{\rm CHSH}=2$, hence  Bell-CHSH inequalities (\ref{bounds}) are not anymore violated. 
\\

\noindent {\bf Objection 1} [from an imaginary quantum physicist]. This is an intriguing example, that makes you think.\footnote{This is roughly the reaction that Bell had, when in the seventies of the last century one of us presented for the first time, at a conference at the CERN, the equivalent `vessels of water model'.} At first, I thought a mistake must be lurking, but the calculations are really very simple and it is easy to check that there is none, and that all reasonings are legitimate. So, the model certainly reveals a mechanism that can be used to violate Bell-CHSH inequalities, but there is little chance that it is the same mechanism in force in quantum entangled micro-systems. Because you see, your model violates Bell-CHSH inequalities with a value of $4$, which is the maximal algebraic violation, and we all know that quantum violations are limited by Cirel'son's bound to the value $2\sqrt{2}$ \citep{cirelson1980}. If only because of this, it seems plausible to me that the quantum situation brings in mechanisms of a very different nature.

\subsection{Variant 2: black or white string}
\label{version 2}

\begin{figure}[!ht]
\centering
\includegraphics[scale =0.24]{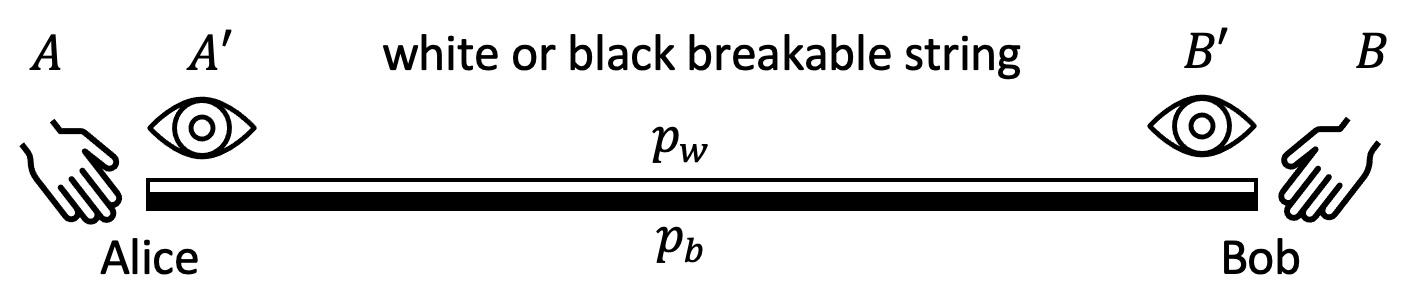}
\caption{Variant 2: a schematic representation of the four experiments $A$, $A'$, $B$ and $B'$ that Alice and Bob can jointly perform on a string of length $L$, which is observed to be white with probability $p_w$ and black with probability $p_b$.}
\label{Version2}
\end{figure}

To address Objection 1, we consider a slightly different situation, so that the magnitude of the violation of Bell-CHSH inequalities can now have arbitrary values, also below Cirel'son's bound. The system on which Alice and Bob perform their joint experiments is always a string of length $L$, made of a breakable material, but this time the color of the string is unstable, in the sense that it randomly oscillates between black and white, with $p_w$ (respectively, $p_b$) the probability that, when observed, it appears as white (respectively, black), with $p_w+p_b=1$. See the schematic representation in Figure~\ref{Version2}. Alternatively, one can think that Alice and Bob's colleague who likes to play pranks, secretly changes the string before each execution of the joint measurements, placing a white string with probability $p_w$ and a black string with probability $p_b$. 

Different form the previous situation, the black color becomes a possible outcome, so we now have the joint probabilities described in Table~\ref{t3}.
\begin{table}[H]
\begin{center}
\begin{tabular}{|
>{\columncolor[HTML]{EFEFEF}}c |c|c|c|c|}
\hline
\cellcolor[HTML]{C0C0C0} & \cellcolor[HTML]{EFEFEF} $P^{++}$ & \cellcolor[HTML]{EFEFEF} $P^{+-}$& \cellcolor[HTML]{EFEFEF} $P^{-+}$& \cellcolor[HTML]{EFEFEF} $P^{--}$\\ \hline
$AB$ & $0$ & $1/2$ & $1/2$ & $0$ \\ \hline
$AB'$ & $p_w$ & $p_b$ & $0$ & $0$ \\ \hline
$A'B$ & $p_w$ & $0$ & $p_b$ & $0$ \\ \hline
$A'B'$ & $p_w$ & $0$ & $0$ & $p_b$ \\ \hline
\end{tabular}
\end{center}
\caption{Variant 2: the $16$ joint probabilities associated with the outcomes of Alice's and Bob's joint measurements on a string which is observed to be white with probability $p_w$ and black with probability $p_b=1-p_w$, producing the values $A_{\rm CHSH}=4p_w$ and $D_{\rm CHSH}=-4(1-p_w)$.\label{t3}}
\end{table}
The situation for joint measurements $AB$ and $A'B'$ is similar to that of Variant 1 of the model, i.e., they produce perfect anticorrelations and perfect correlations, respectively. On the other hand, joint measurements $AB'$ and $A'B$ do not produce anymore perfectly correlated outcomes. More precisely, we have $E_{AB}=-E_{A'B'}=-1$, and $E_{A'B}=E_{AB'}=p_w-p_b=1-2p_b$, so that: $A_{\rm CHSH}=-(-1)+2(1-2p_b)+1=4p_w$, and similar calculations yield: $B_{\rm CHSH}=C_{\rm CHSH}=0$, and $D_{\rm CHSH}=-4p_b$. We see that we can now have a violation of (\ref{bounds}) of arbitrary magnitude. In particular, if we choose $p_w=\sqrt{2}/2$, then $A_{\rm CHSH}=2\sqrt{2}$, which corresponds to Cirel'son's bound. Also, if $p_w=1/2$, then $A_{\rm CHSH}=2$, and $D_{\rm CHSH}=-2$, so we have no violation in this case. 
\\

\noindent {\bf Objection 2}. Interesting to observe that the toy model can violate Bell's inequalities with arbitrary values, so my previous criticism was unfounded. But I have bad news, because I meanwhile discovered that your beautiful model has a flaw: it violates the \emph{no-signaling conditions}, also called \emph{marginal laws}. As you certainly know, these conditions are not violated by the quantum formalism, and this means that the mechanism subtended by your model cannot be the same in force in quantum mechanics. More precisely, if for instance we consider the probability $P_B(A=+)$ that Alice obtains outcome ``$+$,'' when performing experiment $A$, irrespective of the outcome of Bob, when the latter performs measurement $B$, we have: $P_B(A=+)=P_{AB}^{++}+P_{AB}^{+-}=0+1/2=1/2$, but if we calculate the probability $P_{B'}(A=+)$ that Alice obtains outcome ``$+$,'' when performing experiment $A$, irrespective of the outcome of Bob, when the latter performs measurement $B'$, we have: $P_{B'}(A=+)=P_{AB'}^{++}+P_{AB'}^{+-}=1+0=1$. Hence, $P_B(A=+)=1/2\neq 1 =P_{B'}(A=+)$, which is a violation of the marginal laws. And one can of course exhibit many other of these violations in your model.

\subsection{Variant 3: black or white string with length-color correlations}
\label{version 3}

\begin{figure}[!ht]
\centering
\includegraphics[scale =0.24]{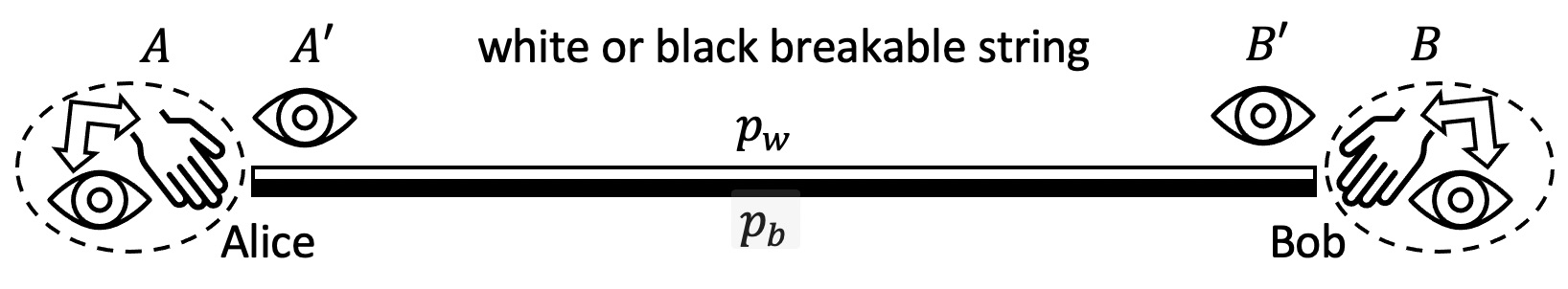}
\caption{Variant 3: a schematic representation of the four experiments $A$, $A'$, $B$ and $B'$ that Alice and Bob can jointly perform on a string of length $L$, which can be white with probability $p_w$ and black with probability $p_b$. Measurements $A$ and $B$, different from previous variants of the experiment, now measure the length-color correlation.}
\label{Version3}
\end{figure}

To address Objection 2, we consider a new variation of the model, such that the degree of violation of the marginal laws can now be varied, hence can also be obeyed. This variant is inspired by \citet{aerts2005}; see also \citep{AertsSassolideBianchi2019}. The system is always a string of length $L$, made of a breakable material whose color oscillates between black and white, with probabilities $p_w$ and $p_b$, respectively. Measurements $A'$ and $B'$ are the same color-measurements as before, but measurements $A$ and $B$ are now different. When Alice executes $A$, she always pulls the string, to measure its length, but she also jointly measures its color, and if the outcomes are ``long-white'' or ``short-black,'' she notes them ``$+$.'' If they are ``long-black'' or ``short-white,'' she notes them ``$-$.'' Here ``long'' means longer than $L/2$, and ``short'' means shorter than $L/2$. In other words, Alice, when performing $A$, now measures the length-color correlation, and Bob does of course the same on his side when performing $B$. See the schematic representation in Figure~\ref{Version3}.

In the calculation of joint probabilities, the difference with respect to Variant 2 of the model is not in the joint measurement $A'B'$, as it is the same, nor in joint measurement $AB$, as is clear that only anticorrelated outcomes can be observed, since Alice and Bob cannot both collect a long fragment, or a short fragment. But there is a difference in the two joint measurements $AB'$ and $A'B$, since now we cannot have anticorrelated outcomes, as is clear that the length is always long, hence we can only have the ``long-white'' and ``long-black'' combinations for the $A$ measurement, when jointly performed with $B'$ (and same for the $B$ measurement, when jointly performed with $A'$). The (long-white,white) situation corresponds to the ``$++$'' outcome and the (long-black,black) situation to the ``$--$'' outcome. Also, the ``$+-$'' outcome, which corresponds to the ``(long-white,black) or (short-black,black)'' situation, and the ``$-+$'' outcome, which corresponds to the ``(short-white,white) or (long-black,white)'' situation, cannot happen, as the string is observed to be either black or white by both Alice and Bob, and never to be shorter than $L/2$. Hence, we have the joint probabilities described in Table~\ref{t4}.
\begin{table}[H]
\begin{center}
\begin{tabular}{|
>{\columncolor[HTML]{EFEFEF}}c |c|c|c|c|}
\hline
\cellcolor[HTML]{C0C0C0} & \cellcolor[HTML]{EFEFEF} $P^{++}$ & \cellcolor[HTML]{EFEFEF} $P^{+-}$& \cellcolor[HTML]{EFEFEF} $P^{-+}$& \cellcolor[HTML]{EFEFEF} $P^{--}$\\ \hline
$AB$ & $0$ & $1/2$ & $1/2$ & $0$ \\ \hline
$AB'$ & $p_w$ & $0$ & $0$ & $p_b$ \\ \hline
$A'B$ & $p_w$ & $0$ & $0$ & $p_b$ \\ \hline
$A'B'$ & $p_w$ & $0$ & $0$ & $p_b$ \\ \hline
\end{tabular}
\end{center}
\caption{Variant 3: the $16$ joint probabilities associated with the outcomes of Alice's and Bob's joint measurements on a string which can be white with probability $p_w$ and black with probability $p_b=1-p_w$, giving the value $A_{\rm CHSH}=4$. Different from Variant 1 and Variant 2 of the model, the $A$ and $B$ experiments now measure the length-color correlation. The marginal laws are obeyed if $p_w=1/2$.\label{t4}}
\end{table}

The main difference with respect to the previous version of the model is that the joint measurements $AB'$ and $A'B$ now produce perfectly correlated outcomes. More precisely, we now have $E_{AB}=-1$ and $E_{A'B'}=E_{A'B}=E_{AB'}=1$, so that $A_{\rm CHSH}=4$, and $B_{\rm CHSH}=C_{\rm CHSH}=0=D_{\rm CHSH}=0$, independently of the value of $p_w$. Considering the marginal laws, and taking again the example of the two marginal probabilities $P_B(A=+)$ and $P_{B'}(A=+)$, we have: $P_B(A=+)=P_{AB}^{++}+P_{AB}^{+-}=0+1/2=1/2$, and $P_{B'}(A=+)=P_{AB'}^{++}+P_{AB'}^{+-}=p_w+0=p_w$. Hence, if we choose $p_w=1/2$, then $P_B(A=+)=P_{B'}(A=+)$, and one can check that all the other marginal relations are also obeyed. 
\\

\noindent {\bf Objection 3}. Congratulations, I'm impressed, you managed to create an experimental situation with a macroscopic entity where you have a violation of the Bell-CHSH inequalities without jointly violating the marginal laws. But it still doesn't convince me, because you see, the price you had to pay to succeed is to have again a fixed algebraically maximum violation of the Bell-CHSH inequalities, hence, I'm now back to my first objection. It seems to me that if you fix one aspect of your model, you get a problem somewhere else, thus confirming my sentiment that it does not capture the real mechanism able to explain a microscopic Bell-test quantum experiment.

\subsection{Variant 4: two black or white strings with length-color correlations}
\label{version 4}

\begin{figure}[!ht]
\centering
\includegraphics[scale =0.24]{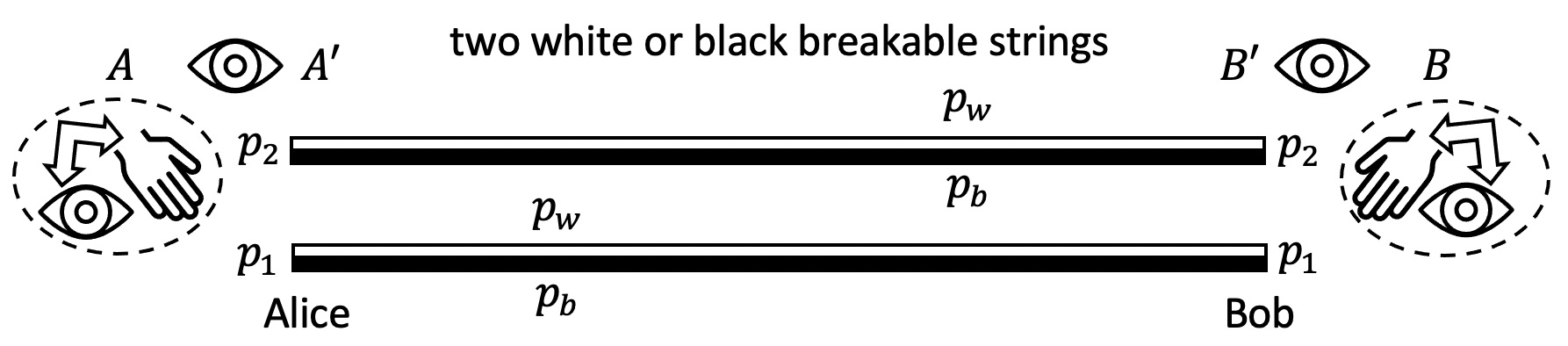}
\caption{Variant 4: a schematic representation of the four experiments $A$, $A'$, $B$ and $B'$ that Alice and Bob can jointly perform on two strings of length $L$, which can be white with probability $p_w$ and black with probability $p_b$. Alice and Bob select string 1 with probability $p_1$ and string 2 with probability $p_2$. Apart from that, measurements are defined as in Variant 3 of the model.}
\label{Version4}
\end{figure}

To address Objection 3, we further elaborate on our model, to obtain a situation where the marginal laws are obeyed and the possibility of varying the magnitude of the violation of the Bell-CHSH inequalities is preserved.\footnote{To the authors' knowledge, a macroscopic model with such properties had never been proposed up to now.} The model no longer consists of a single string, but of two strings. This means that when Alice and Bob perform their measurements, which are the same as those defined in Variant 3 of the model, they have to randomly select one of the two strings, which will be then the one on which they will operate. This means that, different from Variant 3 of the model, there will be situations where Alice and Bob do not select the same string, and situations where they do so. See the schematic representation in Figure~\ref{Version4}.

The calculation of the joint probabilities is now more involved, but presents no difficulties. Let us denote $p_1$ the probability with which Alice and Bob select string 1, and $p_2$ the probability with which they select string 2, with $p_2=1-p_1$. One could of course also consider the situation where these probabilities are different for Alice and Bob, but there is no reason to overcomplicate the model. An important difference with respect to the single string situation is that now Alice can obtain a white (respectively, black) outcome when Bob jointly obtains a black (respectively, white) outcome, for the string color, when they select a different string. Note that when Alice and Bob select a different string, the ``short'' result can never be actualized. With that in mind, let us now explicitly calculate all the joint probabilities (the reader who is not interested in the details can directly jump to the results in Table~\ref{t5}).

Let us start considering $P_{AB}^{++}$. This corresponds to a situation where both Alice and Bob observe either outcome ``long-white,'' or outcome ``short-black''. If they select the same string, the observed colors cannot be different, nor they can collect both a ``short'' or ``long'' string, which means that the ``$++$'' outcome has no contribution from the situations where Alice and Bob select the same string. So, the events contributing to the ``$++$'' outcome, when Alice and Bob select different strings (which are then necessarily ``long''), are the following: ``(Alice selects string 1 \emph{and} is white \emph{and} Bob selects string 2 \emph{and} is white) \emph{or} (Alice selects string 2 \emph{and} is white \emph{and} Bob selects string 2 \emph{and} is white).'' We thus have the joint probability:
\begin{equation}
P_{AB}^{++}=p_1p_wp_2p_w + p_2p_wp_1p_w=2p_1p_2p_w^2
\label{++2}
\end{equation}

By a similar reasoning, the ``$--$'' outcome corresponds to the events: ``(Alice selects string 1 \emph{and} is black \emph{and} Bob selects string 2 \emph{and} is black) \emph{or} (Alice selects string 2 \emph{and} is black \emph{and} Bob selects string 1 \emph{and} is black),'' which gives the joint probability:
\begin{equation}
P_{AB}^{--}=p_1p_bp_2p_b + p_2p_bp_1p_b=2p_1p_2p_b^2
\label{--}
\end{equation}

Let us now consider $P_{AB}^{+-}$. It corresponds to the following events: ``(Alice selects string 1 \emph{and} is white \emph{and} Bob selects string 2 \emph{and} is black) \emph{or} (Alice selects string 2 \emph{and} is white \emph{and} Bob selects string 1 \emph{and} is black) \emph{or} (Alice selects string 1 \emph{and} Bob selects string 1 \emph{and} the string is white \emph{and} Alice's fragment is longer) \emph{or} (Alice selects string 2 \emph{and} Bob selects string 2 \emph{and} the string is white \emph{and} Alice's fragment is longer) \emph{or} (Alice selects string 1 \emph{and} Bob selects string 1 \emph{and} the string is black \emph{and} Alice's fragment is shorter) \emph{or} (Alice selects string 2 \emph{and} Bob selects string 2 \emph{and} the string is black \emph{and} Alice's fragment is shorter).'' This gives the joint probability: 
\begin{eqnarray}
P_{AB}^{+-}&=&p_1p_wp_2p_b + p_2p_wp_1p_b + p_1p_1p_w\frac{1}{2} + p_2p_2p_w\frac{1}{2} + p_1p_1p_b\frac{1}{2} + p_2p_2p_b\frac{1}{2}\nonumber\\
&=&\frac{1}{2}+p_1p_2(2p_wp_b-1)
\end{eqnarray}
and the calculation for $P_{AB}^{-+}$ being specular to that of $P_{AB}^{+-}$, we also have 
\begin{equation}
P_{AB}^{-+}=\frac{1}{2}+p_1p_2(2p_wp_b-1)
\end{equation}

Let us now consider the joint measurement $AB'$, and more precisely the joint probability $P_{AB'}^{++}$. The corresponding events are: ``(Alice selects string 1 \emph{and} is white \emph{and} Bob selects string 2 \emph{and} is white) \emph{or} (Alice selects string 2 \emph{and} is white \emph{and} Bob selects string 1 \emph{and} is white) \emph{or} (Alice selects string 1 \emph{and} is white \emph{and} Bob selects string 1) \emph{or} (Alice selects string 2 \emph{and} is white \emph{and} Bob selects string 2),'' which gives the joint probability: 
\begin{equation}
P_{AB'}^{++}=p_1p_wp_2p_w + p_2p_wp_1p_w + p_1p_1p_w + p_2p_2p_w=p_w(1-2p_1p_2p_b)
\end{equation}

For the joint probability $P_{AB'}^{+-}$, we have the following contributing events: ``(Alice selects string 1 \emph{and} is white \emph{and} Bob selects string 2 \emph{and} is black) \emph{or} (Alice selects string 2 \emph{and} is white \emph{and} Bob selects string 1 \emph{and} is black).'' This gives: 
\begin{equation}
P_{AB'}^{+-}=p_1p_wp_2p_b + p_2p_wp_1p_b = 2p_1p_2p_wp_b
\end{equation}

For the joint probability $P_{AB'}^{--}$, we have to proceed as for probability $P_{AB'}^{++}$, but interchanging the roles of $p_w$ and $p_b$. This gives: 
\begin{equation}
P_{AB'}^{--}=p_b(1-2p_1p_2p_w)
\end{equation}

For the joint measurement $A'B$, we observe that for symmetry reasons we have the same probabilities as for measurement $AB'$. Finally, let us consider the joint measurement $A'B'$. For the calculation of $P_{A'B'}^{++}$, we have to consider the following events: ``(Alice selects string 1 \emph{and} is white \emph{and} Bob selects string 1) \emph{or} (Alice selects string 2 \emph{and} is white \emph{and} Bob selects string 2) \emph{or} (Alice selects string 1 \emph{and} is white \emph{and} Bob selects string 2 \emph{and} is white) \emph{or} (Alice selects string 2 \emph{and} is white \emph{and} Bob selects string 1 \emph{and} is white),'' which gives: 
\begin{equation}
P_{A'B'}^{++}=p_1p_wp_1 + p_2p_wp_2+p_1p_wp_2p_w +p_2p_wp_1p_w=p_w(1-2p_1p_2p_b)
\label{++}
\end{equation}
So, we find that $P_{A'B'}^{++}=P_{AB'}^{++}$, and we leave it to the reader to check that we also have $P_{A'B'}^{+-}=P_{AB'}^{+-}$, $P_{A'B'}^{-+}=P_{AB'}^{-+}$, and $P_{A'B'}^{--}=P_{AB'}^{--}$. Table \ref{t5} summarizes these results.
\begin{table}[H]
\begin{center}
\begin{tabular}{|
>{\columncolor[HTML]{EFEFEF}}c |c|c|c|c|}
\hline
\cellcolor[HTML]{C0C0C0} & \cellcolor[HTML]{EFEFEF} $P^{++}$ & \cellcolor[HTML]{EFEFEF} $P^{+-}$& \cellcolor[HTML]{EFEFEF} $P^{-+}$& \cellcolor[HTML]{EFEFEF} $P^{--}$\\ \hline
$AB$ & $2p_1p_2p_w^2$ & $\frac{1}{2}+p_1p_2(2p_wp_b-1)$ & $\frac{1}{2}+p_1p_2(2p_wp_b-1)$ & $2p_1p_2p_b^2$ \\ \hline
$AB'$ & $p_w(1-2p_1p_2p_b)$ & $2p_1p_2p_wp_b$ & $2p_1p_2p_wp_b$ & $p_b(1-2p_1p_2p_w)$ \\ \hline
$A'B$ & $p_w(1-2p_1p_2p_b)$ & $2p_1p_2p_wp_b$ & $2p_1p_2p_wp_b$ & $p_b(1-2p_1p_2p_w)$ \\ \hline
$A'B'$ & $p_w(1-2p_1p_2p_b)$ & $2p_1p_2p_wp_b$ & $2p_1p_2p_wp_b$ & $p_b(1-2p_1p_2p_w)$ \\ \hline
\end{tabular}
\end{center}
\caption{Variant 4: the $16$ joint probabilities associated with the outcomes of Alice's and Bob's joint measurements on a system formed by two strings which can be white with probability $p_w$ and black with probability $p_b=1-p_w$. Alice and Bob select string 1 with probability $p_1$ and string 2 with probability $p_2=1-p_1$. These probabilities produce the values $A_{\rm CHSH}= 4[1-p_1p_2(1+4p_wp_b)]$ and $B_{\rm CHSH}=C_{\rm CHSH}= D_{\rm CHSH}=4p_1p_2(1-4p_wp_b)$.\label{t5}}
\end{table}

Let us calculate the four correlation functions and the quantities in (\ref{bounds}). We have: 
\begin{eqnarray}
&&E_{AB}=2p_1p_2(p_w^2+p_b^2)-1-2p_1p_2(2p_wp_b-1)=-1+4p_1p_2(1-2p_wp_b)\nonumber\\
&&E_{AB'}=E_{A'B}=E_{A'B'}=1-8p_1p_2p_wp_b
\end{eqnarray}
Thus, for the four combinations (\ref{CHSH-quantity}), we find: 
\begin{eqnarray}
&&A_{\rm CHSH}= 4[1-p_1p_2(1+4p_wp_b)]\nonumber\\ 
&&B_{\rm CHSH}=C_{\rm CHSH}= D_{\rm CHSH}=4p_1p_2(1-4p_wp_b)
\label{CHSH-quantity2}
\end{eqnarray}

Finally, let us also consider the marginal probabilities $P_B(A=+)$ and $P_{B'}(A=+)$. We have: 
\begin{eqnarray}
&&P_B(A=+)=P_{AB}^{++}+P_{AB}^{+-}=\frac{1}{2}+p_1p_2(2p_w^2+2p_wp_b-1)\nonumber\\
&&P_{B'}(A=+)=P_{AB'}^{++}+P_{AB'}^{+-}=p_w
\end{eqnarray}
Hence, the condition 
\begin{equation}
P_B(A=+)-P_{B'}(A=+)=\frac{1}{2}-p_w+p_1p_2(2p_w^2+2p_wp_b-1)=0
\end{equation}
is verified when $p_w=p_b=1/2$, and it is straightorward to check that all the other merginal laws are then also obeyed. Replacing these specific values in (\ref{CHSH-quantity2}), we find: 
\begin{eqnarray}
&&A_{\rm CHSH}= 4(1-2p_1p_2)=4(p_1^2+p_2^2)\nonumber\\ 
&&B_{\rm CHSH}=C_{\rm CHSH}= D_{\rm CHSH}=0
\label{CHSH-quantity3}
\end{eqnarray}

In conclusion, when $p_w=1/2$, the marginal laws are obeyed, and depending on the value of $p_1$, $A_{\rm CHSH}$ can vary from a situation of no violation, with $A_{\rm CHSH}=2$, when $p_1=1/2$, to a situation of an algebraically maximal violation, $A_{\rm CHSH}=4$, when $p_1=0$, or $p_1=1$. Also, when $p_1={1\over 2}(1\pm\sqrt{\sqrt{2}-1})$, $A_{\rm CHSH}=2\sqrt{2}$ is exactly Cirel'son's bound. 
\\

\noindent {\bf Objection 4}. I'm again impressed, you succeeded in responding to all my criticisms. So, I am now forced to admit it, this toy model reveals a mechanism that is general enough to possibly describe quantum entanglement. But I still doubt this is `the' mechanism at work when dealing with micro-systems. Because if that were true, this should somehow emerge also from the quantum formalism, but I do not believe the latter predicts the existence of these mysterious non-local ``hidden'' potential interactions, responsible for the existence of correlations of the second kind.

\section{Measurement interactions}
\label{Measurement interactions}

To address Objection 4, we now abandon the string model and enter the description of the measurement process in the quantum formalism, to emphasize that the latter naturally incorporates the possibility of interpreting quantum probabilities as resulting from ``hidden'' measurement interactions that are contextually actualized at each run of a measurement. 

To see this, let us consider the simple situation of a measurement on a spin-$1/2$ entity, which only has two outcomes. Using Dirac notation, let $\rho=|\psi\rangle\langle\psi|$ be the initial spin state. When we represent it in the Bloch sphere, we can associate to it a 3-dimensional unit real vector ${\bf r}$, such that $\rho={1\over 2}(\mathbb{I} + {\boldsymbol\sigma}\cdot {\bf r})$, where ${\boldsymbol\sigma}$ is a vector formed by the three Pauli matrices, generators of the $SU(2)$ group. 

Let us assume that we are measuring a spin observable $A=P_+-P_-=|+\rangle\langle+|-|-\rangle\langle-|$. In the Bloch representation, there exist two opposite unit vectors, ${\bf n}_\pm$, ${\bf n}_+=-{\bf n}_-,$ such that $P_\pm={1\over 2}(\mathbb{I} + {\boldsymbol\sigma}\cdot {\bf n}_\pm)$. In other words, the measurement of the observable $A$ can be associated with the \emph{one-dimensional diameter-region} subtended by the two outcome unit vectors ${\bf n}_\pm$, which we will call $\triangle_1$.

It is well accepted that a measurement involves a \emph{decoherence process}, consequence of the entanglement resulting from the interaction of the measured entity with the measuring apparatus, producing the disappearance of the non-diagonal elements of the operator state $\rho$, when expressed in the eigenbasis $\{|+\rangle,|-\rangle\}$. This means that the first phase of a measurement can be described as the transition of $\rho$ towards the fully reduced operator state $\rho^\parallel=r_+^\parallel P_+ + r_-^\parallel P_-$, where the coefficients $r_\pm^\parallel = {\rm Tr} \rho P_\pm = |\langle\pm|\psi\rangle|^2$ are the Born probabilities. 

From the perspective of the Bloch sphere, the decoherence process can be described as a movement where the point representative of the state ``dives'' into the sphere, along a rectilinear path orthogonal to the diameter-region $\triangle_1$, until it stops exactly on that region. In other words, we have a first phase in the measurement corresponding to the transition $\rho\to\rho^\parallel$, where one can show that $\rho^\parallel={1\over 2}(\mathbb{I} + {\boldsymbol\sigma}\cdot {\bf r}^\parallel)$, with ${\bf r}^\parallel=r_+^\parallel {\bf n}_+ + r_-^\parallel {\bf n}_-$, the vector ${\bf r}^\parallel$ being parallel to $\triangle_1$, which explains our use of the $\parallel$-symbol.

It is at this point that the formalism reveals the possible existence of hidden (potential) interactions. Indeed, the vector $\rho^\parallel$ splits the region $\triangle_1$ in a way that is exactly proportional to the outcome probabilities, in the sense that the sub-region $A_+$, going from ${\bf n}_-$ to ${\bf r}^\parallel$, has length $\mu(A_+)=2r_+^\parallel$, and the sub-region $A_-$, going from ${\bf r}^\parallel$ to ${\bf n}_+$, has length $\mu(A_-)=2r_-^\parallel$. And since $\mu(\triangle_1)=2$, one observes that the ratios $r_\pm^\parallel = \mu(A_\pm)/\mu(\triangle_1)$ are exactly the Born probabilities. This means that one can interpret $\triangle_1$ as a \emph{potentiality region} describing all the available interactions between the measuring and measured systems, with those in $A_+$ producing the collapse ${\bf r}^\parallel\to {\bf n}_+$, and those in $A_-$ producing the collapse ${\bf r}^\parallel\to {\bf n}_-$. 

\begin{figure}[!ht]
\centering
\includegraphics[scale =0.3]{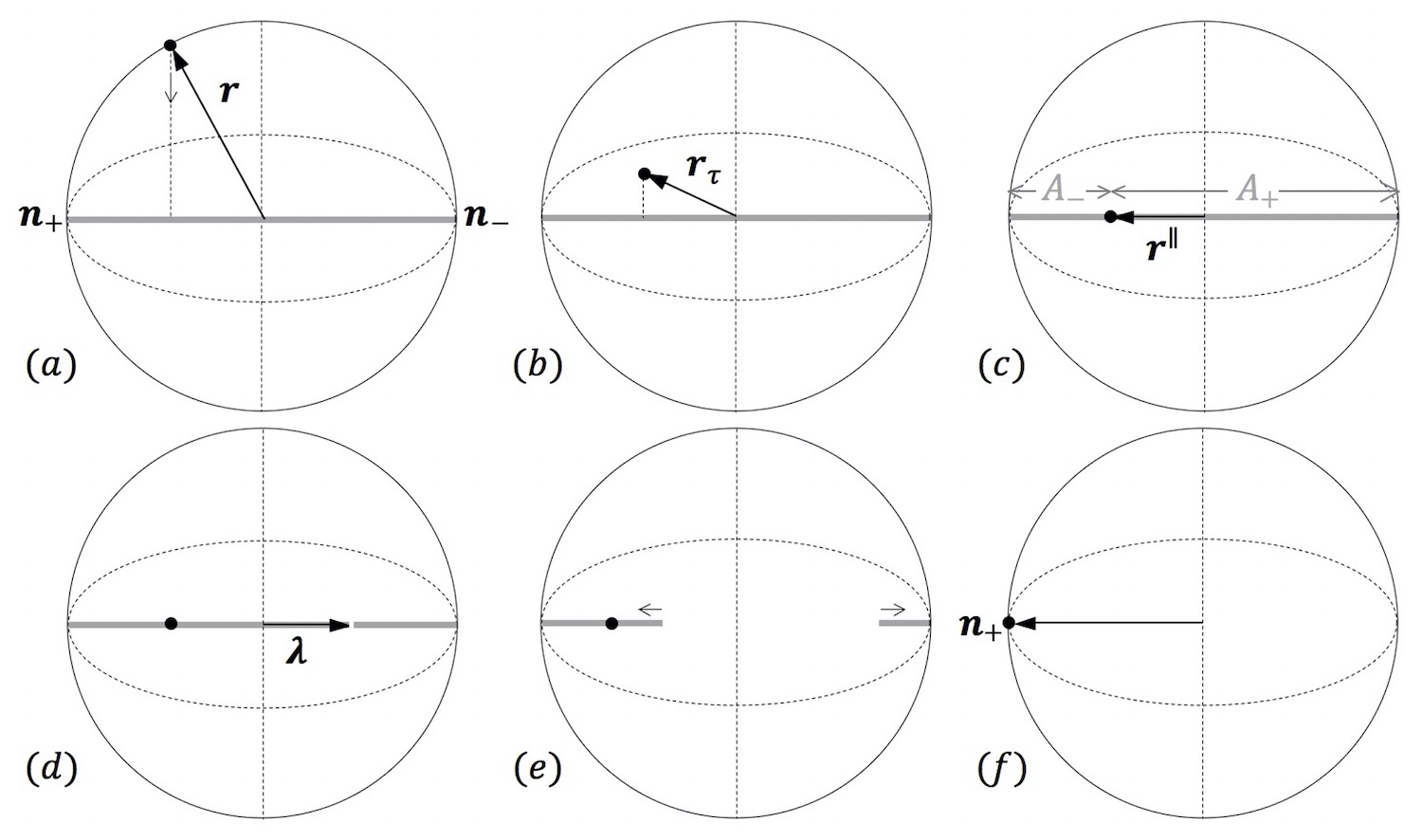}
\caption{The unfolding of a two-outcome measurement process within the Bloch sphere. First, there is a decoherence process, where the state dives into the Bloch sphere, a movement which can be parametrized as: ${\bf r}_\tau=(1-\tau) {\bf r} + \tau {\bf r}^\parallel$, $\tau\in[0,1]$, with ${\bf r}_0={\bf r}$, and ${\bf r}_1={\bf r}^\parallel$. Viewing the fully reduced decohered state ${\bf r}^\parallel$ as an abstract point particle sticked on a uniform elastic band, corresponding to the potentiality region $\triangle_1$ (a 1-dimensional simplex), when the band breaks at some unpredictable point, here point $\boldsymbol\lambda\in A_+$, by contracting, it brings the particle to one of its two end points, here ${\bf n}_+$, completing the measurement process.}
\label{Collapse}
\end{figure}

There is an interesting way to visualize this process. One can imagine the potentiality region to be like an abstract elastic band that can break at some unpredictable point, which by collapsing pulls the point particle representative of the state towards one of the two outcome states, according to the projection postulate. The hidden interactions are then in a correspondence with all these potential breaking points of the elastic band; see Figure~\ref{Collapse}. 

It is of course not the scope of this paper to delve into the mathematics of the \emph{hidden measurement interpretation of quantum mechanics}, and we refer the interested readers to \citet{AertsSassolideBianchi2014,AertsSassoli2016} for more details. What is important here to emphasize is that the above description is not limited to 2-dimensional systems and can be generalized to measurements having an arbitrary number $N$ of outcomes, possibly also degenerate. Then, the potentiality region, describing the hidden interactions, becomes a $(N-1)$-dimensional simplex $\triangle_N$, inscribed in the convex region of states belonging to a generalized $(N^2-1)$-dimensional Bloch sphere \citep{Arvind1997, Kimura2003, Byrd2003, Kimura2005, Bengtsson2006, Bengtsson2013}. These simplexes can always be viewed as abstract hyper-membranes that the pre-measurement state ${\bf r}$, following the decoherence process ${\bf r}\to {\bf r}^\parallel$, partitions into $N$ convex sub-regions, and one can prove that the relative sizes of these regions correspond exactly to the quantum probabilities \citep{AertsSassolideBianchi2014}. 
\\

\noindent {\bf Objection 5}. I wasn't aware that the Bloch representation could be generalized to an arbitrary number of dimensions $N$ of the state space, and that the eigenvectors that characterize an observable determine in it a simplex-shaped region, that the decohered state is capable of partitioning into as many sub-regions as there are outcomes of the measurement, and that the relative sizes of these sub-regions are exactly the probabilities predicted by Born's rule. This certainly makes it more credible that there could be an objective ``weighted symmetry breaking mechanism,'' based on hidden interactions, at the origin of the quantum collapses and associated probabilities, in the hypothesis of course that such collapses would be objective physical phenomena and not mere Bayesian updates of an experimenter's knowledge. But let me express a last objection. Coming back to your string model, it is obvious that quantum correlations, if they are truly created by the joint measurements, they must originate from a connective element that somehow unites the two sub-entities. In your model, this connective element is the very body of the string. But when considering two micro-entities, like two entangled electrons, or two entangled photons, we know that when they are separated by arbitrarily large distances there is nothing that can be detected in the space between them, playing the role of such hypothetical connective element. Ultimately, I believe this is the fundamental reason why your interesting string model is unable to capture the gist of what happens during a typical Bell-test experiment with micro-entities.

\section{Concluding remarks}

This last objection brings us closer to the true mystery of quantum entanglement, and more generally of quantum superposition states. Again, we want to present some arguments that will help clarify why this mechanism of correlations of the second kind is able to properly illustrate what happens with the entanglement phenomenon. Here we will do so from a more general perspective of the foundations and philosophy of physics. 

Let us first emphasize that the reason we consider simple mechanical models of breaking strings as examples relevant to the phenomenon of entanglement is not in the attempt to give quantum mechanics a more classical interpretation. We believe that the unquestionable success of quantum mechanics as a predictive physical theory shows that the microscopic reality has a profoundly different nature from the everyday reality that surrounds us, thus from the classical mechanic reality. So, why these string models to illustrate the phenomenon of entanglement? 

To answer this question, we must consider another one: ``What are the properties of physical reality that are fundamentally different in quantum reality from classical reality?" This is what is really important to consider first, and in doing so, it is appropriate to be cautious in any of our attempts to interpret those micro-world phenomena that confront us with a ``quantum singularity." With this in mind, it is 
also important to remember that the phenomenon of quantum entanglement is primarily linked to the superposition principle, an entangled state being in general a superposition of product states. But the superposition principle is generally valid in quantum mechanics, even in the situation of a system formed by a single entity. This observation is relevant because there are strong reasons to believe that the absence of a connective element in space between two entangled entities participating in a Bell-test experiment, unlike the situation in our string models, can be explained by a general property of quantum entities that is also possessed by single quantum entities.

Indeed, since the early days of quantum mechanics, there were strong suspicions that the Schr\"{o}dinger wave function should not be regarded as a wave present in space, but as an expression of the potentialities by which a quantum entity can be located in certain regions of space. In pondering this possibility, of particular importance were the experiments conducted in the 1970s by Helmut Rauch, with low-energy neutrons \citep{rauchetal1974,rauchetal1975}, when one of us was a doctoral student in the quantum physics group at the University of Geneva, where the work of Rauch's experimental group had attracted much attention.\footnote{Interestingly, Anton Zeilinger, who was recently awarded the Nobel Prize for his experimental work on entanglement, co-authored Rauch's main experiment \citep{rauchetal1975}, as he was one of his students at the time.} 

In fact, these astonishing experiments made it particularly clear that it was not possible to think of the wave function as a wave spreading in space, in the sense that such hypothesis was insufficient to explain the non-local effects that these experiments revealed, which led to the introduction, by one of the authors, of the notion of \emph{non-spatiality}, already from the early 1990s \citep{Aerts1990,aerts1998}. The other author of this article was also working in Geneva, a decade later, when interest in neutron interferometry experiments was still very much alive. This partly explains why their collaboration around the notion of non-spatiality could naturally emerge, also influencing other fields of their joint research \citep{sassolidebianchi2017,sassolidebianchi2021}. 

More exactly, it was possible to deduce from these experiments that a neutron can show its joint presence in two separate spatial regions, within an interferometer, with nothing being present in the space between them. Despite this separation, the neutron continues to behave as a whole entity, thus demonstrating that in the quantum layer of reality wholeness and interconnectedness do not depend on having actual spatial connections between the different elements that form an entity. 

We believe that what happens with the interconnectedness of entangled entities must find an explanation that is similar to what happens to the neutrons in Rauch's experiments. It is also important to note that, in the meantime, similar experiments have been carried out that can also bring complex entities like molecules, consisting of more than $800$ atoms, in such  kind of non-local superposition state \citep{gerlichetal2011}, showing that they can remain whole entities throughout an experiment despite manifesting in different spatial regions, with nothing in between them. It is such behavior with respect to space, of a single entity, or of two entangled entities, that we call non-spatiality. 

Compatibly with the notion of non-spatiality, we believe that the non-linear and indeterministic evolution of the wave function, usually referred to as \emph{collapse}, or \emph{reduction}, is to be considered as the more general process of change, while the linear and deterministic one, described by the Schr\"{o}dinger equation, should be considered as the special case, applicable only when there is no interaction with a measuring apparatus. This is the way quantum physics is regarded today by most physicists who use it, but for some reason it is rarely the way it is regarded by those who reflect on its philosophical interpretation. 

This means that the philosophical maneuver leading to the Many Worlds interpretation \citep{Everett1957,DeWitt1973} would not be necessary, in the sense that the indeterministic evolution, which perhaps for the time being is described too simply on the basis of the collapse of the wave function, should not disappear and be absorbed in the Schr\"{o}dinger evolution, but in a sense exactly the opposite \citep{AertsSassolideBianchi2015}.

In the interpretive approach we have proposed with our collaborators, it is also important that the fundamental evolution in the quantum world remains indeterministic, to allow both physics and psychology, when more advanced, to integrate within a single explanatory framework the free choice of a person who decides whether and which experiment to carry out. In our approach, steps are already taken in that direction, since the indeterminism of a quantum measurement is described to occur due to the presence of fluctuations on the interactions between the measuring apparatus and the entity being measured. 

Note that a form of this indeterminism also exists in classical mechanics. Indeed, if a classical entity is in a state of unstable equilibrium, it is the fluctuations present in the perturbations of this state that push the classical entity out of it, and these fluctuations are usually unpredictable, i.e., random in nature. We then speak of \emph{bifurcations}. The difference with the quantum situation is that the set of unstable states in a classical theory always has measure zero, and in this sense the uncertainties associated with it do not contribute to the probabilities associated with outcomes of the measurements. Superposition states in quantum mechanics, on the contrary, always form the largest part of the set of all states, so that the indeterminism they carry does contribute to the probabilities of the outcomes of a measurement. 

The presence of an irreducible uncertainty in quantum mechanics, but also in classical mechanics if we consider the unstable equilibrium states, does not mean that the entities being measured, the measuring apparatuses, the experimenters and their environments, cannot proceed as a whole according to a deterministic evolution. So, in principle, what is now called \emph{superdeterminism} \citep{Brans1988,Sabine2020} is a hypothesis that can be put forward as possibly true. However, what the adherents of such an interpretive view seem not to be aware of, is how important to our most basic notion of reality, when viewed as an operational construction, is that we possess a genuine free choice, in the sense of being able to select our experiences among a set of possibilities that are truly such. 

This possibility, of having been able to make different choices in our past, is what really determines the content and richness of our present reality, and is at the very basis of \emph{operationalism}, which in turn is at the core of the project of science. But it would take us too far to analyze this statement and its scope in detail here, so we invite the interested reader to \citet{ Aerts1996,aertssassolidebianchi2023}. Given the main topic of this article, it is also fitting to note that the hypothesis of the existence of free choice is crucial in terms of interpreting the meaning of a Bell-test experiment, when Bell inequalities are violated. Indeed, if experimenters are not assumed to be able to freely choose the orientations of the polarizers in measurements on polarization-entangled photon pairs, the very significance of the design of these experiments is lost \citep{scheidletal2010}. 

Coming back to the notion of non-spatiality, it is worth observing that a widespread ``classical'' prejudice persists today, consequence of the way in which we humans have experienced reality during our evolution on this planet, which is to believe that our physical world is somehow fully contained in space, and more generally in spacetime. As a consequence, as we observed already, if two entities are separated by a very large spatial distance, they are also believed to be experimentally separated, when we act on them simultaneously. This prejudice is really what led to the initial disbelief about the quantum entanglement phenomenon, but experimental data, via the violation of Bell-CHSH inequalities, told us a very different story, that we could no longer ignore. But we were then also left with the uncomfortable feeling that the entanglement phenomenon, however real, was impossible to explain. 

On the other hand, if space and time are considered to be emergent, i.e., coming into being with the formation of macroscopic aggregates of matter, the unjustified assumption would be to consider that two entities are necessarily disconnected if we cannot detect anything measurable in the space between them, functioning as a connective element. Indeed, a connective element may well exist without being detectable as a specific spatial element. The \emph{extended Bloch formalism}, which we briefly mentioned in Section~\ref{Measurement interactions}, also suggests the existence of such non-spatial missing piece `in between' two entangled entities. The formalism actually does much more than this, as it also provides a less rudimentary description of the non-linear, quantum collapse, indeterministic processes of change, than the standard formalism allows. 

To be more specific, take the case of two spin-$1/2$ entities. Since the Hilbert space is $4$-dimensional, the generalized Bloch sphere is $15$-dimensional, and a state $\rho=|\psi\rangle\langle \psi|$ can always be written as 
\begin{equation}
\rho = {1\over 4}(\mathbb{I} +\sqrt{6}\, {\bf r}\cdot\mbox{\boldmath$\Lambda$})
\end{equation}
where $\boldsymbol\Lambda$ is a 15-dimensional vector formed by a determination of the generators of $SU(4)$. We can take them to be \citep{AertsSassoli2016}: $\Lambda_1={1\over \sqrt{2}} \sigma_1\otimes{\mathbb I}$, $\Lambda_2= {1\over \sqrt{2}}\sigma_2\otimes{\mathbb I}$, $\Lambda_3={1\over \sqrt{2}} \sigma_3\otimes{\mathbb I}$, $\Lambda_4={1\over \sqrt{2}}{\mathbb I}\otimes\sigma_1$, $\Lambda_5={1\over \sqrt{2}}{\mathbb I}\otimes\sigma_2$, $\Lambda_6={1\over \sqrt{2}}{\mathbb I}\otimes\sigma_3$, $\Lambda_7={1\over \sqrt{2}}\sigma_1\otimes\sigma_1$, \dots, $\Lambda_{15}={1\over \sqrt{2}}\sigma_3\otimes\sigma_3$. By direct calculation, one can then show that the vector ${\bf r}$ associated with the state $\rho$, in the Blochean representation, can always be written as the \emph{direct sum}:
\begin{equation}
{\bf r} = {1\over \sqrt{3}}\, {\bf r}_{\rm Alice}\oplus {1\over \sqrt{3}}\, {\bf r}_{\rm Bob}\oplus {\bf r}_{\rm conn},
\label{direct sum}
\end{equation}
where ${\bf r}_{\rm Alice}$ and $ {\bf r}_{\rm Bob}$ are the $3$-dimensional Bloch vectors describing the states of Alice's and Bob's sub-systems, respectively, and ${\bf r}_{\rm conn}$ is a $9$-dimensional vector describing the ``state of their connection.'' 

When the composite system is in a product state, then ${\bf r}_{\rm conn}$ is trivial, in the sense of being fully determined by the components of ${\bf r}_{\rm Alice}$ and $ {\bf r}_{\rm Bob}$. However, when the composite system is in an entangled state, ${\bf r}_{\rm conn}$ cannot anymore be deduced from the knowledge of the sub-systems' states ${\bf r}_{\rm Alice}$ and $ {\bf r}_{\rm Bob}$, as it now describes a genuine additional element of reality, in accordance with the principle that the whole can be greater than the sum of its parts. The important observation for our discussion is that ${\bf r}_{\rm conn}$ is higher dimensional compared to ${\bf r}_{\rm Alice}$ and $ {\bf r}_{\rm Bob}$, which suggests that the entanglement connection would belong to a more abstract layer of our reality, less prone to be detectable by our three-dimensional measuring instruments. 

Regarding the general notion of non-linear and indeterministic change discussed above, we believe that at least some of this change does not occur within space, parameterized by time, thus within spacetime. This means that the so-called `collapse models' of the measurement process, which are usually proposed with the aim of making quantum theory classical again, will not lead to satisfactory results, because they carry with them the bias that change must be a process that occurs within spacetime. In other words, more sophisticated models, also from a philosophical point of view, will be necessary to take into due account the ingredient of non-spatiality. 

Note also that in the string models we have described, there is the simplifying assumption that all points of a string have equal probability to become a breaking point. In the extended Bloch formalism, this assumption of uniform probability over the different interactions is what allows to derive the Born probabilities. But one can relax such condition and also consider situations where some interactions are more probable than others, i.e., where the probability distributions are not anymore uniform. In this more general framework, the `uniform case' is situated somehow in between two limit cases, the `classical case', where almost all outcomes are predetermined once the measured entity is in a given state, and the `solipsistic case', where only the fluctuations are responsible for the (non-predeterminable) outcome, with no influence from the state \citep{aertsetal1993,aertsdurt1994,aertsetal1997a,aertsetal1997b}. 

In that sense, one can say that if a quantum measurement is described by the standard collapse model in Hilbert space, then it lies in between a situation of pure discovery (of what exists prior to the measurement) and pure creation (of something that did not exist prior to the measurement). However, this feature of the pure quantum measurements cannot be captured by remaining within the Hilbert space formalism. It is necessary for this to consider a more fine grained approach, like the one of the extended Bloch formalism, where probability distributions are not restricted to be uniform. The Born probabilities can then be shown to appear as \emph{universal averages}, i.e., as averages over all different possible probability distributions, describing all possible ways of selecting a measurement-interaction  \citep{AertsSassolideBianchi2014}. 

This possibility, of viewing quantum measurements as \emph{universal measurements}, i.e., as averages over different typologies of measurements, classical, solipsistic and in between, holds a possible explanation of the classicality of the macroscopic material world. Indeed, this would be due to the presence of an oversupply of classical-like measurements in the collection of those on which the average operates. Does it mean that non-classical, or even solipsistic measurements, would be totally absent in the macroscopic material reality? Surely not, as the examples of measurements with the strings we proposed in this article clearly demonstrates. Simply, these typologies of measurements are not usually considered necessary to gain relevant knowledge about macroscopic material entities, like strings, so in particular they are not used in Bell-test experimental protocols. 

One fascinating aspect of the hypothesis that quantum measurements are to be equated with universal measurements, is that it allows free choice 
to be described in a way that is compatible with quantum mechanics and the general measurement model we propose, which generalizes the quantum model. Indeed, to give an example, free choice dictated by criteria of rationality, or by a moral code, since it doesn't contemplate every possible ``way of choosing," but only a tiny sub-class of such ways, it cannot be associated with Born probabilities, but with something more close to a classical probabilistic model. On the other hand, persons who make their choices in ways that we would define as irrational, not being governed by specific principles, hence more open to actualize every possible ``way of choosing," would be more closely described by a quantum model. Also, the fluctuations that give rise to the bifurcations present in classical unstable systems, when viewed as part of measurements, they are to be associated with solipsistic-like processes, and corresponding statistics, in the classification scheme we have introduced. 

Coming now to the question of \emph{marginal laws}, we have seen that our model allows for their violation, something that is not possible in quantum mechanics, if one uses the same tensor product representation to describe all the joint measurements as product measurements. However, the violation is not an essential ingredient of it, in the sense that Bell-CHSH inequalities can be easily violated, with arbitrary magnitude, also when the marginal laws are obeyed. This possibility of violating the marginal laws in our model, and in other models built with similar logic \citep{AertsSassolideBianchi2021}, opens up the possibility that what is predicted by the standard quantum formalism may not necessarily be always the rule. And in that respect, we can observe that in experiments the marginal laws are in fact violated, although these violations are usually attributed to experimental errors \citep{AdenierKhrennikov2007,DeRaedt2012,DeRaedt2013,AdenierKhrennikov2016,Bednorz2017,Kupczynski2017}.

If these violations truly happen, i.e., are not just experimental errors, it means that the entanglement phenomenon should not only be associated with the states, but also with the joint measurements. Indeed, when the marginal laws and Bell-CHSH inequalities are jointly violated, to model the data one needs to also consider non-product measurements (if one wants to have the same pre-measurement state for each joint measurement), which means that joint measurements should also be considered to be entangled. 

Quoting from \citet{aertsetal2019}: ``Thinking of entanglement as being present also at the level of the measurements might seem like a very drastic perspective [...], particularly in those experimental situations where there is a clear spatial separation between the measurement apparatuses working in a coincident way. However, if the measured entity forms a whole, it is to be expected that also the measurements can become entangled, precisely through the very wholeness of the measured entity, because their action on the latter would occur simultaneously and not sequentially. In other words, the notions of locality and separability, usually intended as `spatial locality' and `spatial separability', need here be replaced by the more general notions of `sub-system locality' and `sub-system separability'. This because among the salient properties of physical and conceptual systems, there is precisely that of non-spatiality, and therefore `separation in space' is not anymore a sufficient criterion for characterizing a separation of two sub-systems and corresponding joint measurements.'' 

We think it is important to point out that it is not just entanglement that reveals to us that our physical reality is mostly non-spatial. \emph{Quantum superposition}, \emph{quantum measurement}, \emph{quantum complementarity} and \emph{quantum indistinguishability} are all phenomena that remain unintelligible if we hold on to our prejudice that microphysical entities are permanent resident of our spatiotemporal theater. 

Without going into details here, let us mention an interpretation called the \emph{conceptuality interpretation}, that we are currently further developing in our group \citep{aertsetal2020}. In its framework, the nature of non-spatiality is addressed in a very natural way, by considering that physical micro-entities are to be interpreted as ``entities of meaning," which therefore can be found in more or less abstract states, with respect to a given semantic context. Non-spatiality would then be an expression of \emph{conceptual abstractness}, and this opens up a truly novel perspective on the physical world and how quantum mechanics should be interpreted, although, of course, such perspective remains mostly still a subject for further study for the time being.

Much more should be said to fully clarify the scope of the models we have presented and how they fit into the study of more general physical entities, 
but this would require much more space than what is available here. To conclude, we have presented different variants of a simple string model that can explain entanglement as being the result of two ingredients: (1) potential measurement interactions between the measured system and the measuring apparatus; (2) an element of reality connecting the two sub-systems from which correlations of the second kind are produced. 

In our string model, the existence of the potential measurement interactions presents no mysteries, as they clearly correspond to the different ways Alice and Bob can jointly pull the string, producing different breaking points, hence different correlations. The presence of the connective element is also non-mysterious, as the unbroken string can be interpreted as an entity formed by two string fragments that are not in a well-defined length-state, in the same way the two spin-$1/2$ entities in a singlet state cannot be associated with well-defined spatial directions. In particular, there is no ``spooky action at a distance'' between Alice and Bob, and no superluminal communication between them. 

We have argued that our macroscopic model remains relevant also for the microscopic domain. Indeed, micro-entities can remain whole even when they appear fragmented from a spatial perspective. This is a general quantum property that holds not only for entangled entities, but also for individual entities. It is usually referred to as non-locality, but should be more properly understood as non-spatiality, or non-spatiotemporality, and is a fundamental feature of all microscopic quantum systems.

\end{document}